**Giant Magnetoelectric Effect in Antiferromagnetic $BaMnO_{3-\delta}$ and Its Derivatives**

O.B. Korneta[1,2], T. F. Qi[1,2], M. Ge[1,2,3], S. Parkin[1,4], L.E. DeLong[1,2], P. Schlottmann[5], and G. Cao[1,2*]

[1]Center for Advanced Materials, University of Kentucky, Lexington, KY 40506

[2]Department of Physics and Astronomy, University of Kentucky, Lexington, KY 40506

[3]Department of Physics, University of Science and Technology of China, Hefei, China

[4]Department of Chemistry, University of Kentucky, Lexington, KY 40506

[5]Department of Physics, Florida State University, Tallahassee, FL 32306

Hexagonal perovskite 15R-$BaMnO_{2.99}$ with a ratio of cubic to hexagonal layers of 1/5 in the unit cell is an antiferromagnetic insulator that orders at a Néel temperature $T_N$ = 220 K. Here we report structural, magnetic, dielectric and thermal properties of single crystal $BaMnO_{2.99}$ and its derivatives $BaMn_{0.97}Li_{0.03}O_3$ and $Ba_{0.97}K_{0.03}MnO_3$. The central findings of this work are: **(1)** these materials possess a usually large, high-temperature magnetoelectric effect that amplifies the dielectric constant by more than an order of magnitude near their respective Néel temperature; **(2)** Li and K doping can readily vary the ratio of cubic to hexagonal layers and cause drastic changes in dielectric and magnetic properties; in particular, a mere 3% Li substitution for Mn significantly weakens the magnetic anisotropy and relaxes the lattice; consequently, the dielectric constant for both the a- and c-axis sharply rises to 2500 near the Néel temperature. This lattice softening is also accompanied by weak polarization. These findings provide a new paradigm for developing *novel, high-temperature magnetoelectric materials* that may eventually contribute to technology.

## Introduction

Magnetoelectric (ME) material**s** are of fundamental and technological importance and have attracted renewed interest **[1-7]**, including an ongoing quest for new ME materials exhibiting high transition temperatures (e.g., room temperature) to both (anti)ferroelectric and (anti)ferromagnetic states **[4-16]**. While new materials have emerged **[6-16]** in recent years, remarkably few exhibit coexisting electrically or magnetically ordered states below distinct transition temperatures that simultaneously exceed 100 K **[17-21]**. The scarcity of such materials is commonly attributed to a fundamental incompatibility of magnetic and electric order. In the case of $ABO_3$ perovskite materials, ferroelectricity is favored when nonmagnetic $d^0$ transition metal ions occupy the B site, whereas magnetic order requires magnetic (non-$d^0$) ions on the same site. The prevailing view is that such materials are more likely to be found in situations where antiferroelecticity and ferromagnetism are driven independently **[6,21]**.

Recently, first-principles studies have predicted that a ferroelectric ground state driven by the off-centering of a magnetic $Mn^{4+}$ ($d^3$) ion can be artificially established via strain or chemical manipulation in the cubic perovskite $CaMnO_3$ **[22]** and the hypothetical cubic perovskite $BaMnO_3$ **[23]**. The naturally occurring $BaMnO_3$ phase exists only in the hexagonal perovskite structure because the Goldschmidt tolerance factor is greater than unity, i.e., the ionic radius of the $Ba^{2+}$ ion is too large to accommodate the ideal cubic perovskite structure with Pm3m symmetry. It is important to note that the lattice parameters, *a*, *c* and *V* of the ideal cubic (c) and hexagonal (h) perovskite structures of composition $ABO_3$ are simply related: $a_h =$ $a_c$, $c_h$ $=$ $a_c$ and $V_h=3V_c$. Moreover, intermediate between the ideal hexagonal and cubic perovskite structures are structures with different proportions of both cubic layers (corner-sharing octahedra) and hexagonal layers (face-sharing octahedra); examples are the $BaMnO_{3-\delta}$ ($\delta < 0.09$) polytypes, in which the ratio of cubic to hexagonal layers, or the cubic stacking ratio, sensitively depends on chemical composition and/or oxygen content **[24**, **26]**. The close relationship between cubic and hexagonal perovskite structures, along with an easily tuned cubic stacking ratio, makes $BaMnO_3$ a unique archetype for studying the correlation between structural and physical properties in general, and spin-lattice coupling and ME behavior in particular. In light of the renewed interest in $BaMnO_3$, this study is particularly timely and necessary given the fact that research on the physical properties of $BaMnO_3$ is remarkably scanty in literature compared to that on the cubic perovskite manganites and hexagonal mangnites $RMnO_3$ (R = Y and smaller rare earths) although the structural studies of the polytypes have been much more thoroughly investigated **[25-31]**.

Here we report a systematic study on structural, magnetic, dielectric and thermal properties of single-crystal $BaMnO_{2.99}$ and its derivatives $BaMn_{0.97}Li_{0.03}O_3$ and $Ba_{0.97}K_{0.03}MnO_3$, as a function of the cubic stacking ratio that varies between 1/6, 1/5 and 1/4. This study reveals that these materials have a usually large, high-temperature ME effect that is characterized by a more than 10-fold enhancement in the dielectric constant ε(T) near their respective Neel temperatures. Moreover, the highly anisotropic dielectric response of $BaMnO_{2.99}$, is closely associated with the magnetic anisotropy; in particular, a reduced magnetic anisotropy due to an increase in the cubic stacking ratio leads to lattice softening and a drastic rise in ε(T) for electric field applied along either the **a-** or **c-**axes near the Néel temperature, which is also accompanied by weak polarization. These findings open a new avenue for developing *high-temperature magnetoelectric* materials that may eventually contribute to technology.

## Experimental

Single-crystals of compositions $BaMnO_{3-\delta}$, $BaMn_{0.97}Li_{0.03}O_3$, and $Ba_{0.97}K_{0.03}MnO_3$ were grown at ambient pressure using flux techniques **[24].** The structures of these single-crystal samples were refined at both 90 and 295 K using a Nonius-Kappa CCD single-crystal X-ray diffractometer (XRD) and the SHELX-97 programs **[24]**. Chemical compositions were checked using energy dispersive X-ray (EDX) analyses. Measurements of specific heat C(T,H) and magnetization M(T,H) were performed using either a Quantum Design PPMS or Quantum Design MPMS. The complex permittivity $\varepsilon(T,H,\omega) = \varepsilon' + i\varepsilon''$ was measured using a QuadTech Model 7600 LCR Meter with a frequency range of 10 Hz $\leq \omega \leq$ 2 MHz in applied magnetic fields $\mu_o H \leq 9$ T. The electric polarization P was measured over a temperature range $1.8 \leq T \leq 400$ K and applied fields $\mu_o H \leq 9$ T using a Radiant Precision Premier II polarimeter. The resistivities of all single crystals exceeded $10^8$ $\Omega$ cm below 400 K, and electrical leakage was therefore negligible.

## Results and Discussion

Single-crystal samples of $BaMnO_{3-\delta}$ were found to adopt the "15R phase" with space group R-3m (#166) and a cubic stacking ratio of 1/5, as shown in **Fig. 1**. This structure perfectly matches that previously reported for $BaMnO_{2.99}$ **[26]**. $Ba_{0.97}K_{0.03}MnO_3$ and $BaMn_{0.97}Li_{0.03}O_3$ form in the 12H and 8H phases with space group $P6_3/mmc$ (#194), respectively; the corresponding cubic stacking ratio is 1/6 for the former and ¼ for the latter (see **Fig. 1** and **Table 1)**. A close examination of the refined structures for the doped $BaMnO_3$ phases reveals a strong preference for substituted species to locate within the hexagonal layers rather than within the cubic layers **[22]**. This tendency suggests the bonding interactions are relatively strong within the cubic layers, consistent with earlier studies **[25, 26, 29]**. Moreover, the Mn ions in the cubic and hexagonal layers are asymmetrically situated within the $MnO_6$ octahedra with an apparent tendency for Mn ions to shift positions toward the nearest neighboring cubic layers, resulting in uneven Mn-O and Mn-Mn bond distances **[24]**.

**Table 1. Sample Structural Data at 295 K [90 K]**

| Compound | Cubic stacking | a (Å) [90K] | c (Å) [90K] | Space group |
|---|---|---|---|---|
| **15R-$BaMnO_{2.99}$** | 1/5 | 5.6772(1) [5.6662(8)] | 35.3314(10) [35.276(7)] | R-3m |
| **12H-$Ba_{0.97}K_{0.03}MnO_3$** | 1/6 | 5.6832(10) [5.673(1)] | 28.3760(6) [28.3395(7)] | $P6_3/mmc$ |
| **8H-$BaMn_{0.97}Li_{0.03}O_3$** | 1/4 | 5.6734(1) [5.6619(8)] | 18.7496(5) [18.721(4)] | $P6_3/mmc$ |

For a $Mn^{4+}$ ($3d^3$) ion in $BaMnO_{2.99}$, it is expected that all degeneracies are lifted, and the three orbitals with lowest energy are half-filled and form a spin 3/2 for each $Mn^{4+}$ site. The half-occupied orbitals correspond to the $t_{2g}$ states in cubic symmetry. The basal-plane magnetic susceptibility $\chi_a(T)$ (**Fig. 2a)** indicates an antiferromagnetic state with a Néel temperature $T_N$ = 220 K; remarkably, no anomaly is discerned in the **c-**axis $\chi_c(T)$, indicating that the magnetic anisotropy is strong in $BaMnO_{2.99}$. A corresponding lambda anomaly in the specific heat C(T) (**Fig. 2a**, right scale) confirms $T_N$ to be a 2$^{nd}$-order phase transition. It is noted that both $\chi_a(T)$ and $\chi_c(T)$ exhibit extremely weak, non-Curie-Weiss temperature dependence for $T > T_N$; in fact, $\chi_a(T)$ even increases slightly with increasing temperature, implying that there may exist an additional magnetic state at higher temperatures. Below $T_N$, the Mn spins clearly lie within the basal plane for the following two reasons: An anomaly in $\chi_a(T)$ is seen at $T_N$ but not in $\chi_c(T)$, and between $T_N$ and T = 43 K (where another magnetic anomaly occurs), $\chi_a(T)$ decreases with temperature, whereas $\chi_c(T)$ remains essentially temperature-independent. The sharp upturn in both $\chi_a(T)$ and $\chi_c(T)$ at T = 43 K may be attributed to a gradual canting of spin that is also suggested by a broad peak in dC/dT close to 40 K (see in **Fig2a Inset**). This point is also consistent with results of a neutron scattering study on polycrystalline $BaMnO_{2.99}$ **[26]**. The isothermal magnetization below $T_N$ exhibits a linear field-dependence for applied fields up to 7 T, as anticipated for an antiferromagnetic state.

One of the central findings in this study is that magnetic behavior is strongly coupled with dielectric response in $BaMnO_{2.99}$, as illustrated in **Figs. 2a** and **2b**. This is particularly evident in data for $\chi_a$(T), C(T) and the **a**-axis dielectric constant $\varepsilon_a'$(T), which exhibit simultaneous anomalies near $T_N$ = 220 K ($\varepsilon_a'$(T) displays a shoulder near $T_N$, and a more prominent peak at a higher temperature $T_E$ = 273 K). Three major features emerge in **Fig. 2**: **(1)** $\varepsilon_a'$(T) is one order of magnitude greater than $\varepsilon_c'$(T) near $T_N$, consistent with a magnetic response by spins aligned within the basal plane (**Fig. 2a**). (**2)** Application of magnetic field H of 9 T suppresses $T_E$ by 11 K and the main peak magnitude of $\varepsilon_a'$(T) by ~ 13% (defined as [$\varepsilon_a'$($T_E$,0)- $\varepsilon_a'$($T_E$,9T)]/ $\varepsilon_a'$($T_E$,0)) (**Fig. 2b**). $\varepsilon_c'$(T) exhibits a very weak magnetic field dependence, and is not shown here. **(3)** The magnitude of $\varepsilon_a'$(T) --- particularly near $T_N$ and $T_E$ --- is surprisingly large and comparable to, or even greater than, that exhibited by some well-known magnetoelectrics such as $BaMnF_4$ **[2]**, $BiMnO_3$ **[10]**, $HoMnO_3$ and $YMnO_3$ **[13]**.

Dielectric relaxation in $BaMnO_{2.99}$ is evident at very low frequencies (f < 50 kHz), as shown in **Fig. 2c**. Note that the peak in $\varepsilon_a'$(T), which is sharply located at $T_E$ = 273 K for f = 100 Hz, gets broadened for frequencies ***both above and below*** 100 Hz, and also shifts upward in temperature with increasing frequency. The data of $\varepsilon_a'$(T) for f = 20 Hz is consistent with a Curie-Weiss law for 270 K < T < 365 K, and a corresponding fit yields a Curie-Weiss temperature of 257 K (see **Fig. 2c**, right scale), which is intermediate between $T_N$ = 220 K and $T_E$

= 273 K. $\varepsilon_c'$(T) is also frequency-dependent but develops an additional peak at higher temperatures (**Fig. 2d**). It is worth mentioning that the observed frequency dependence of $\varepsilon'$(T) is typical of a relaxor rather than a ferroelectric; but on the other hand, $\varepsilon_a'$(T) obeys a Curie-Weiss law (**Fig.2 inset**), a characteristic consistent with a typical ferroelectric but absent in a relaxor. In addition, the observed low-frequency dielectric response is remarkably non-linear near $T_E$ and mimics the behavior of certain ferroelectrics such as $AgNa(NO_2)_2$, $Pb_3MgNb_2O_9$, etc., that are primarily driven by a soft mode whose frequency vanishes at the phase transition **[31]**. It is remarkable that the frequency dependence of the peak temperature in $\varepsilon_a'$(T) follows the Arrhenius law as shown in **Fig.3**, indicating gaps and hence activation behavior. The activation energy is of the order of 0.45 eV, which is too large to be phononic; therefore it must be an electronic activation. The attempt frequency is 10 GHz, i.e. a microwave frequency; this frequency is much smaller than the activation energy, which is expected as attempt frequencies are usually small compared to the activation gap.

Furthermore, the strong coupling of the dielectric constants $\varepsilon_a'$(T) and $\varepsilon_c$'(T) with the magnetic susceptibility $\chi_a$(T) and $\chi_c$(T) must be attributed to a spin-lattice coupling that displaces the Mn ions relative to the anions when magnetic ordering occurs. It is already noted that the crystal structure of $BaMnO_3$ is unusually sensitive to slight changes in chemical composition, such as slight variations of oxygen content **[26]** and Li and K doping (discussed below); this sensitivity implies that the lattice of $BaMnO_3$ is extraordinarily "soft", therefore susceptible to even slight displacements of the Mn ions due to magnetic ordering.

Indeed, the spin-lattice coupling in $BaMnO_{2.99}$ is so strong that merely a 3% Li substitution for Mn can drastically change both the magnetic and dielectric behavior. As shown in **Fig. 1** and **Table 1**, a 3% Li doping increases the fraction of cubic layers within the unit cell, raising the cubic stacking ratio from 1/5 to 1/4. This structural change generates more pathways along $180^o$ Mn-O-Mn bond angles between the corner-sharing $MnO_6$ octahedra along the **c-**axis, and should be accompanied by a more isotropic magnetic ordering. As shown in **Fig. 4a**, both $\chi_a(T)$ and $\chi_c(T)$ display a strong magnetic anomaly, in sharp contrast to the magnetic behavior for undoped $BaMnO_{2.99}$, where $\chi_c(T)$ shows no discernable magnetic anomaly at $T_N$ (**Fig. 2a**). It is interesting that the magnetic anomaly in $\chi_a(T)$ and $\chi_c(T)$ occurs at different temperatures: $T_{Na}$ = 215 K in the former and $T_{Nc}$ = 290 K in the latter susceptibility (**Fig. 4a).** We also note that C(T) displays an anomaly at 247 K, approximately midway between $T_{Na}$ and $T_{Nc}$ (see **Fig. 4a**, right scale).

The temperature dependences of $\varepsilon_a'(T)$ and $\varepsilon_c'(T)$ closely follow those of $\chi_a(T)$ and $\chi_c(T)$ by peaking at $T_{Ea} \approx T_{Na}$ = 215 K, and $T_{Ec} \approx T_{Nc}$ = 290 K, respectively (see the shaded areas in **Fig. 4b)**. The peak magnitudes of both $\varepsilon_a'(T)$ and $\varepsilon_c'(T)$ approach 2500, which is much larger than that of $BaMnO_{2.99}$. In particular, $\varepsilon_c'(T)$ for $BaMn_{0.97}Li_{0.03}O_3$ at $T \approx T_{Ec}$ = 290 K is more than one order of magnitude greater than that for $BaMnO_{2.99}$ at $T_E$ = 273 K. This radical enhancement in $\varepsilon_c'(T)$ is a signature of a significant lattice softening that is directly associated with a reduced magnetic anisotropy due to the increased cubic stacking ratio. The inferred strong spin-lattice coupling is further corroborated by the magnetic field response of both $\varepsilon_a'(T)$ and $\varepsilon_c'(T)$, with the latter having a stronger field-dependence (**Figs. 5a** and **5b**). The field-dependence of $\varepsilon_c'(T)$ increases with decreasing frequency as the magneto-dielectric ratio (defined as $[\varepsilon_c'(9T) - \varepsilon_c'(0)]/ \varepsilon_c'(0)$ (%)) rises from 20% for 1 kHz to 100 % for 20 Hz, (see **Fig. 5c**).

The above observations concerning the marked differences in magneto-dielectric behavior between $BaMn_{0.97}Li_{0.03}O_3$ and $BaMnO_{2.99}$ strongly suggest that only a modest 3% Li doping has significant impact on spin-lattice coupling via the increased cubic stacking ratio. It is therefore not surprising that this trend is reversed when the cubic stacking ratio is reduced, as can be observed in $Ba_{0.97}K_{0.03}MnO_3$, where the ratio is reduced from 1/5 for $BaMnO_{2.99}$ to 1/6 with only a 3% K doping (**Fig. 1** and **Table 1**). As shown in **Fig. 6a**, $\chi_a(T)$ signals magnetic ordering at $T_N$ = 230 K, whereas no corresponding magnetic anomaly is discerned in $\chi_c(T)$; this behavior is indicative of a stronger magnetic anisotropy due to a weakened superexchange interaction along the **c**-axis. ***This stronger magnetic anisotropy, in turn, hardens the lattice, leading to a weaker dielectric response, particularly, along the c-axis*** (see **Fig. 6b**). It is also noted that $\varepsilon_a^{'}(T)$ exhibits a slope change at $T_N$ = 230 K, but peaks at a much higher temperature $T_E$ = 305 K.

The near-simultaneous anomalies in both the magnetic susceptibility and the dielectric constant in these materials reveal that the dielectric response is critically linked to the antiferromagnetic order. Specifically, the spin-lattice coupling displaces the Mn ions, and changes the phonon mode when the magnetic ordering occurs. Such changes could be very subtle, but sufficient to trigger a radical dielectric response or a phase transition **[33]**, as illustrated in **Fig. 7**, which compares $\varepsilon_a^{'}(T)$ and $\varepsilon_c^{'}(T)$ for all three compounds studied. One feature of **Fig. 7** is particularly remarkable: ***the dielectric response is enhanced by increasing cubic stacking ratio.***

A recent study using first-principles density-functional theory predicts a ferroelectric ground state in a hypothetical cubic peroviskite $BaMnO_3$ **[23]**. In essence, the $t_{2g}$ and $e_g$ states of the magnetic $Mn^{4+}$ ions are strongly hybridized with the O-2p orbitals; this hybridization shifts a delicate balance between competing energies governing the second-order Jahn-Teller effect that favors off-centering, thus a ferroelectric instability **[23]**. (The second-order Jahn-Teller effect is associated with the relaxation of the electronic system in response to the ionic displacements through covalent bond formation **[23]**)  This prediction is further corroborated by anomalously large Born effective charges **[23]** that describe a rearrangement of electrons through covalent-bond formation. Anomalously large Born effective charges (larger than the formal charges on the ions) may indicate off-centering distortions or a ferroelectric instability that is characterized by a large dielectric constant.  Here, the observation of a large dielectric constant that increases with increasing cubic stacking ratio suggests a favorable tendency toward a ferroelectric instability. Indeed, a close examination of the crystal structures of all three compositions under study demonstrate that the Mn-Mn bond distances between hexagonal layers and between cubic and hexagonal layers increase systematically with cubic stacking ratio (see **Fig. 7c**).  This increased Mn-Mn bond distance weakens the magnetic anisotropy, and through spin-lattice coupling, relaxes the lattice. This explains the parallel increase in both $\varepsilon_a'(T)$ and $\varepsilon_c'(T)$ near $T_N$, which eventually reach 2500 for $BaMn_{0.97}Li_{0.03}O_3$ (**Figs. 4** and **7**); in addition, $BaMn_{0.97}Li_{0.03}O_3$ also exhibits net electric polarization and hysteresis consistent with ferroelectric order, as shown in **Fig. 8**.

In summary, $BaMnO_{3-\delta}$ and lightly Li- and K-doped $BaMnO_{3-\delta}$ possess strong magnetoelectric effects near room temperature.  The large dielectric constants of these materials, comparable to those of well-known ferroelectrics, are closely associated with magnetic order and magnetic anisotropy that sensitively hinges on the cubic stacking ratio.  These findings provide ***a new paradigm for developing novel, high-temperature magnetoelectrics and, eventually, multiferroic materials*** of technological importance.

This work was supported by NSF through grants DMR-0552267, DMR-0856234 (GC) and EPS-0814194 (GC, LED), and by DoE through grants DE-FG02-97ER45653 (LED) and DE-FG02-98ER45707 (PS)

***Corresponding author; email: cao@uky.edu**

**References**

1. E. Ascher, H. Reider, H. Schmid and H. Stossel J. Appl. Phys. 37, 1404 (1966)
2. J.F. Scott, Phys. Rev. B **16** 2329 (1977)
3. H. Schmid, Ferroelectrics **162**, 665 (1994)
4. W. Eerenstein, N.D. Mathur and J.F. Scott, Nature **442**, 759 (2006)
5. N.A. Spaldin, S.-W. Cheong, and R. Ramesh, Physics Today, October 2010, p.38
6. D.I. Khomskii, J. Mag. Mag. Mat. **306**, 1 (2006)
7. A. N. Hill, J. Phys. Chem. B **104**, 6694 (2000)
8. Alessio Filippetti, Nicola A. Hill, J. Mag. Mag. Mat. **236** 176 (2001)
9. M. Fiebig, Th Lottermoser, D. Frohlich, A.V. Goltsev and R.V. Pisarev, Nature **419**, 818 (2002)
10. T. Kimura, S. Kawamoto, I. Yamda, M. Azuma, M.Takano and Tokura, Phys. Rev B **67,** 180401 (R) (2003)
11. T. Kimura, et al, Nature **426**, 55 (2003)
12. Bas B. Van Aken, Thomas T. M. Palstra, Alessio Filippetti and Nicola A. Spaldin, Nature Materials **3**, 164 (2004)
13. B. Lorenz, Y. Q. Wang, Y.Y. Sun and C.W. Chu, Phys. Rev. B **70**, 212412 (2004)
14. B. Lorenz, A.P. Litvinchuk, M.M. Gospodinov, and C.W. Chu, Phys. Rev Lett. **92**, 087204 (2004)
15. N. Hur, S. Park, P. A. Sharma, J. S. Ahn, S. Guha and S-W. Cheong, Nature **429**, 392 (2004)
16. R. Ranjith, A. K. Kunu, M. Filippi, B. Kundys. W. Prellier, B. Ravaeu J. Lavediere, M.P. Singh and S. Jandl, Appl. Phys. Lett. **92**, 062909 (2008)
17. G. Cao, J. W. O'Reilly, J. E. Crow and L. R. Testardi, Phys. Rev. B **47**, 11510 (1993)

18. M. Fiebig, J. Phys. D 38 R123 (2005)

19. Z. J. Huang, Y.Cao, Y.Y. Sun, Y.Y. Xue and C.W. Chu, Phys. Rev B **56**, 2623 (1997)

20. Th. Lonkai, D.G. Tomuta, U. Amann, J. Ihringer, R.W. A. Hendrikx, D. M. Tobben and J.A. Mydosh, Phys. Rev. B **69**, 134108 (2004)

21. R.R. Ramesh and N. Spaldin, Nature Mater. **6** 21 (2007)

22. S. Bhattacharjee, E. Bousquet and P. Ghosez, Phys. Rev. Lett **102**, 117602 (2009)

23. J. M. Rondinelli, A.S. Eidelson, and N. A. Spaldin, Phys. Rev. B **79**, 205119 (2009)

24. T.F. Qi, S. Parkin, and G. Cao to be published elsewhere

25. T. Negas and R.S. Roth, J. Solid State Chem. **3**, 323 (1971)

26. Joesephine J. Adkin and Michael A. Haywards, Chem. Mater. **19**, 755 (2007)

27. L. Katz and R. Ward, Inorg. Chem. **3**, 205 (1964)

28. B.L. Chamberland, A.W. Sleight and J.F. Weiher, J. Solid State Chem. **1**, 506 (1970)

29. A.J. Jacobson and A. J.W. Horrox, Acta Crystallogr., Sect. B **32**, 1003 (1976)

30. E.J. Cussen and P.D. Battle, Chem. Mater. **12**, 831 (2000)

31. Joesephine J. Adkin and Michael A. Haywards, J. Solid State Chem **179** 70(2006)

32. E.E. Lines and A.M. Glass, *Principles and Applications of Ferroelectrics and Related Materials* (Clarendon Press, Oxford, 1977), pages 140, 286, 393.

33. D. Frohlich, St. Leute, V.V. Pavlov, R. V. Pisarev, Phys. Rev. Lett **81** 3239 (1998)

**Captions**

**Fig.1.** Stacking sequences and cubic stacking ratios based on our single crystal x-ray diffraction for $Ba_{0.97}K_{0.03}MnO_3$, $BaMnO_{2.99}$, and $BaMn_{0.97}Li_{0.03}O_3$ (from left to right).

**Fig.2.** The temperature dependence for $BaMnO_{2.99}$ of **(a)** magnetic susceptibility $\chi_a(T)$ for the a-axis and $\chi_c(T)$ for the c-axis at $\mu_oH$ = 0.1 T, and C(T) (right scale); **(b)** dielectric constant $\varepsilon_a'(T)$ for the a-axis and $\varepsilon_c'(T)$ for the c-axis at frequency $\omega$=100 Hz and $\mu_oH$ = 0 T (solid line) and 9 T (dashed line); **(c)** $\varepsilon_a'(T)$ at a few representative frequencies and a fit for the Curie-Weiss law $1/\varepsilon_a'(T)$ (right scale); and **(d)** $\varepsilon_c'(T)$ at a few representative frequencies. Inset: dC/dT vs T.

**Fig.3**. Logarithmic frequency, ln(f), vs. reciprocal peak temperature, 1/T, of $\varepsilon_a'(T)$ for $BaMnO_{2.99}$.

**Fig.4.** The temperature dependence for $BaMn_{0.97}Li_{0.03}O_3$ of **(a)** magnetic susceptibility $\chi_a(T)$ for the a-axis and $\chi_c(T)$ for the c-axis at $\mu_oH$ = 0.1 T, and C(T) (right scale); **(b)** dielectric constant $\varepsilon_a'(T)$ for the a-axis and $\varepsilon_c'(T)$ for the c-axis at frequency $\omega$=100 Hz.

**Fig.5.** The temperature dependence for $BaMn_{0.97}Li_{0.03}O_3$ of **(a)** $\varepsilon_a'(T)$ at a few representative frequencies; **(b)** $\varepsilon_c'(T)$ at a few representative frequencies at $\mu_oH$ = 0 T (solid lines) and 9 T (dashed lines), and **(c)** the magneto-dielectric ratio, defined as $[\varepsilon_c'(9T) - \varepsilon_c'(0)]/ \varepsilon_c'(0)$ (%).

**Fig.6.** The temperature dependence for $Ba_{0.97}K_{0.03}MnO_3$ of **(a)** magnetic susceptibility $\chi_a(T)$ for the a-axis and $\chi_c(T)$ for the c-axis at $\mu_oH$ = 0.1 T; **(b)** dielectric constant $\varepsilon_a'(T)$ for the a-axis and $\varepsilon_c'(T)$ for the c-axis at frequency $\omega$=100 Hz.

**Fig.7.** Comparisons of **(a)** $\varepsilon_a'(T)$ and **(b)** $\varepsilon_c'(T)$ at frequency $\omega$=100 Hz, and **(c)** the Mn-Mn bond distances between hexagonal layers and between cubic and hexagonal layers (right scale) as a function of the cubic stacking ratio for $BaMnO_{2.99}$, $BaMn_{0.97}Li_{0.03}O_3$ and $Ba_{0.97}K_{0.03}MnO_3$.

**Fig.8.** The electric polarization at T = 150 K for $BaMn_{0.97}Li_{0.03}O_3$.

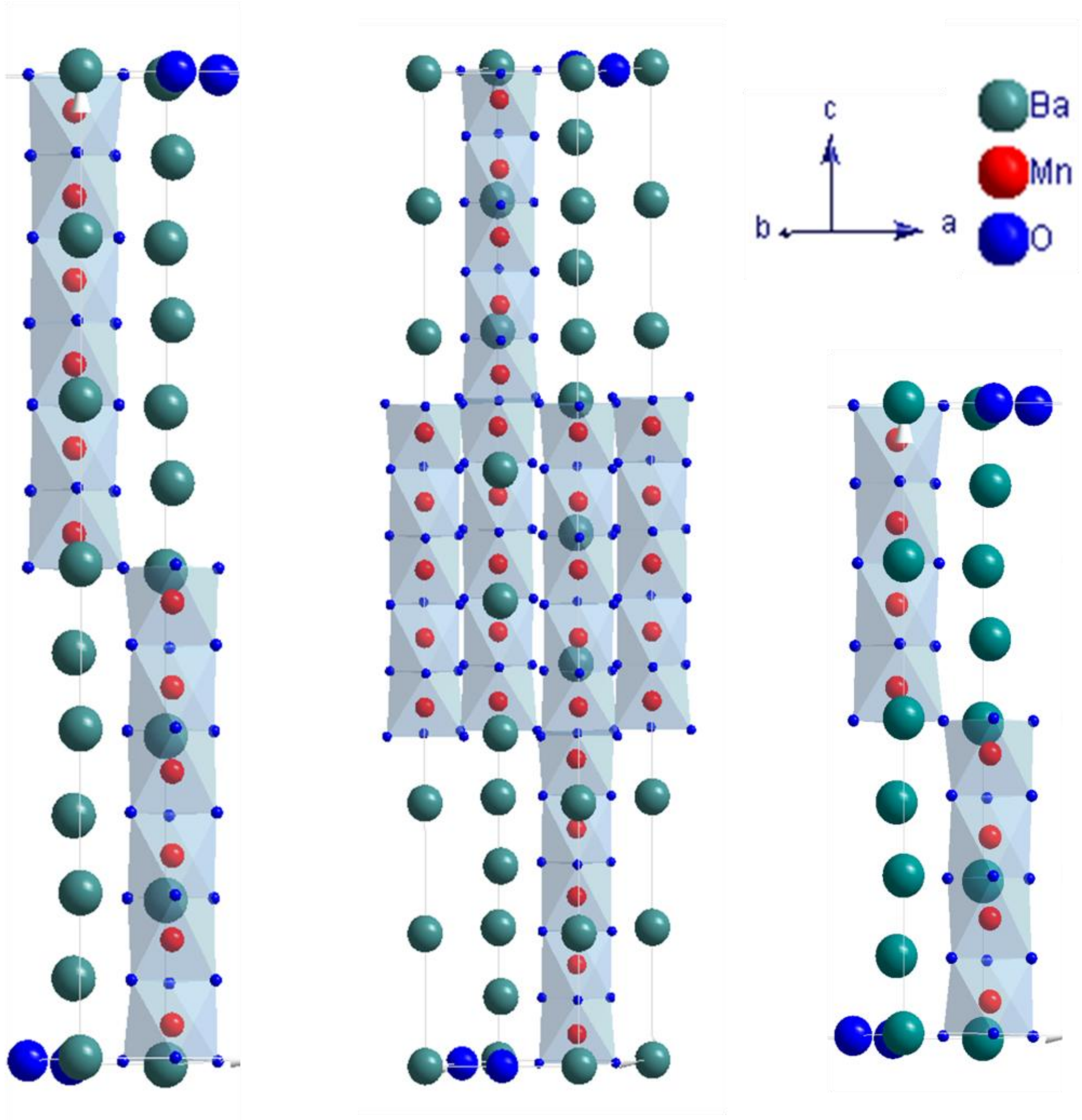

**12H-$Ba_{0.97}K_{0.03}MnO_3$** Cubic stacking: **1/6**

**15R-$BaMnO_{2.99}$** Cubic stacking:**1/5**

**8H-$BaMn_{0.97}Li_{0.03}O_3$** Cubic stacking:**1/4**

Fig.1

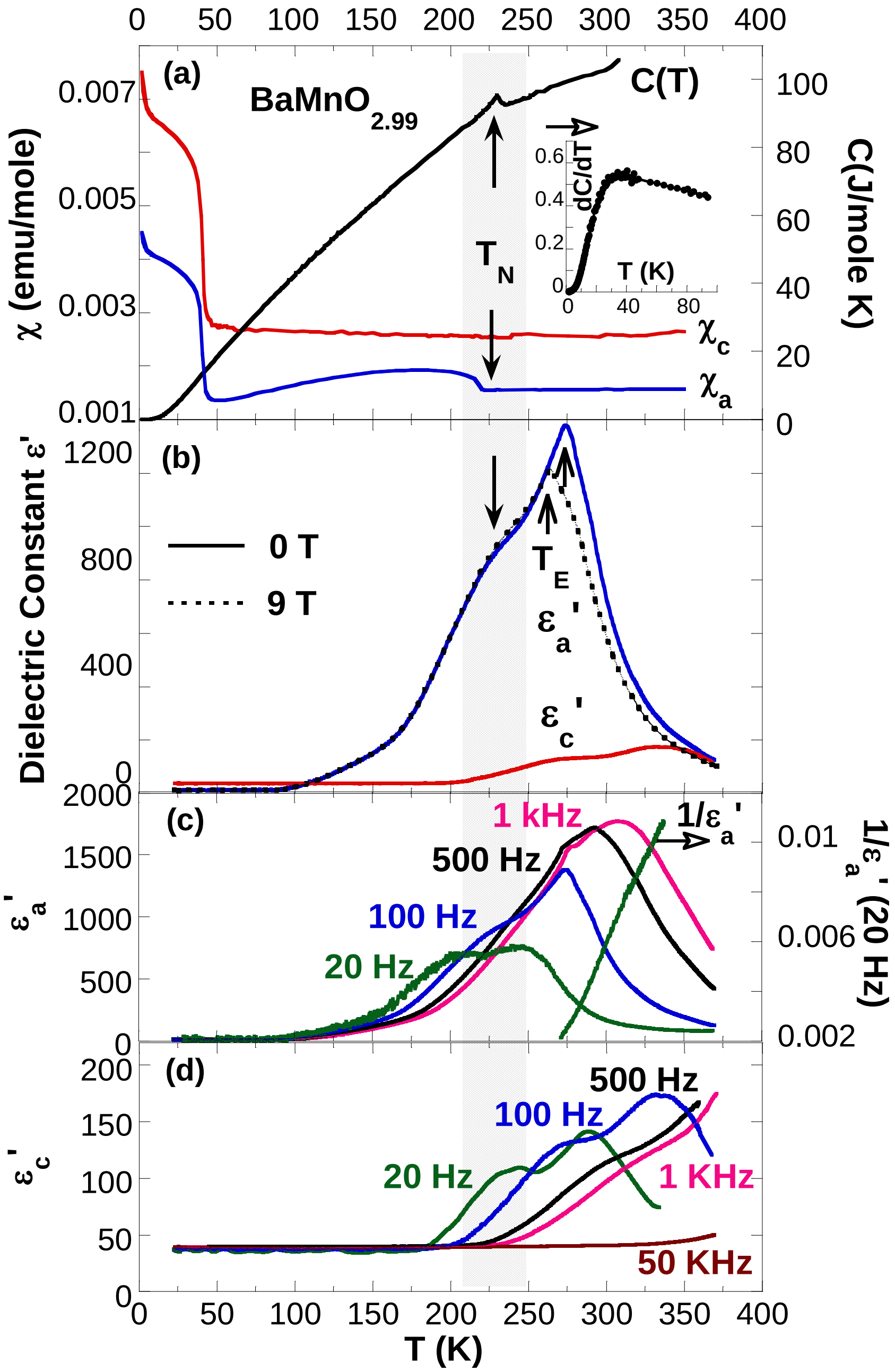

Fig.2

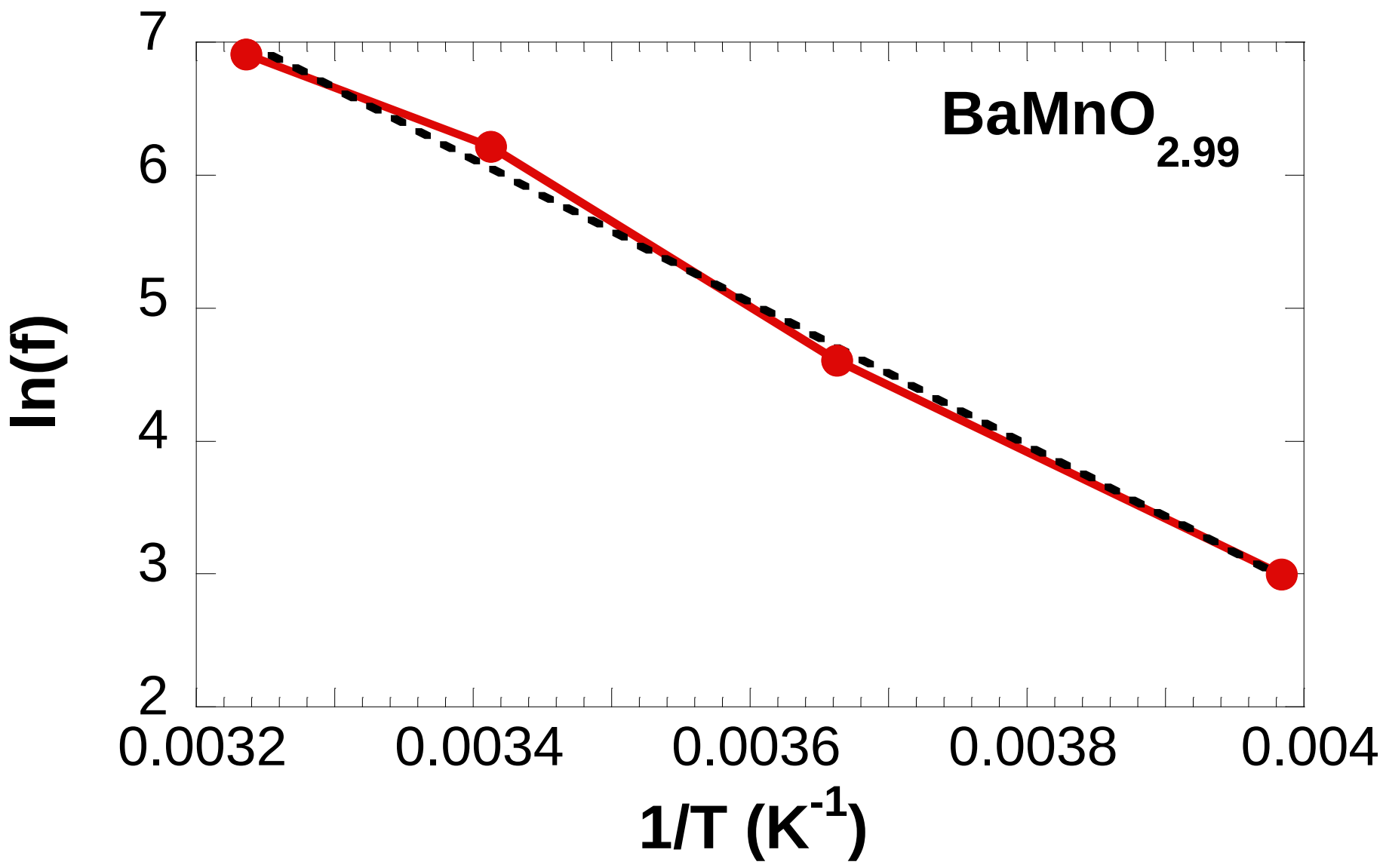

Fig.3

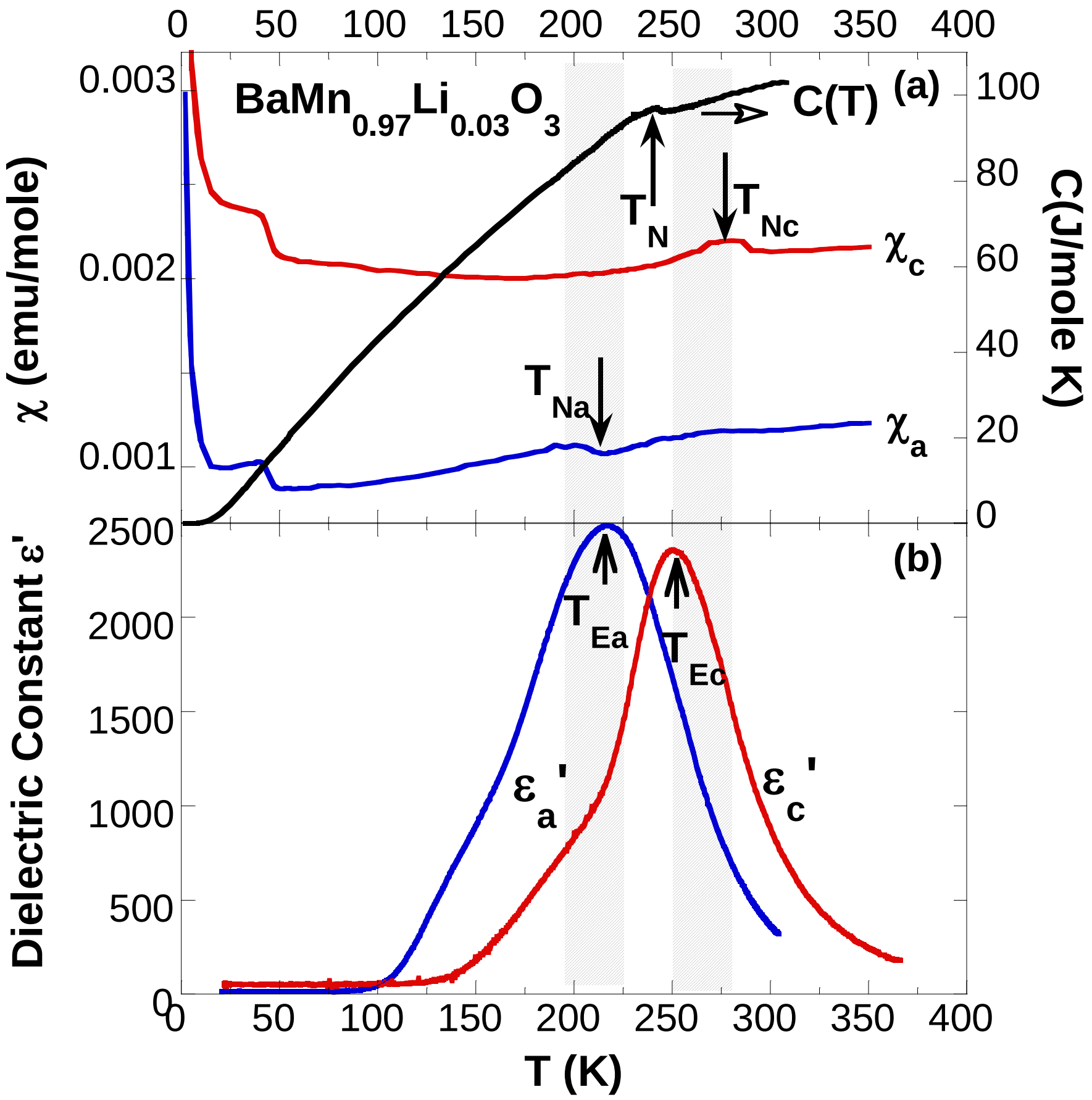

Fig. 4

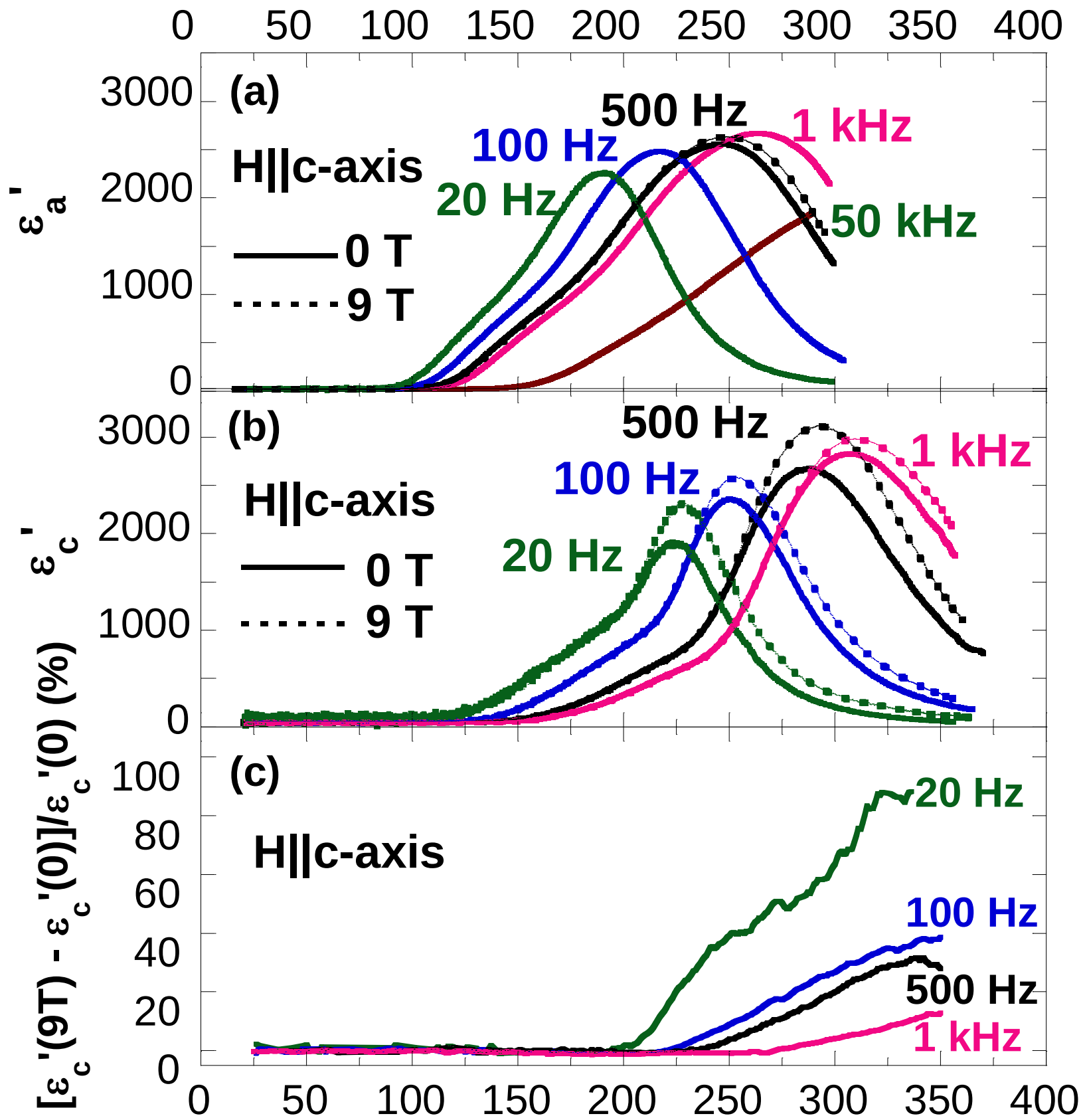

Fig.5

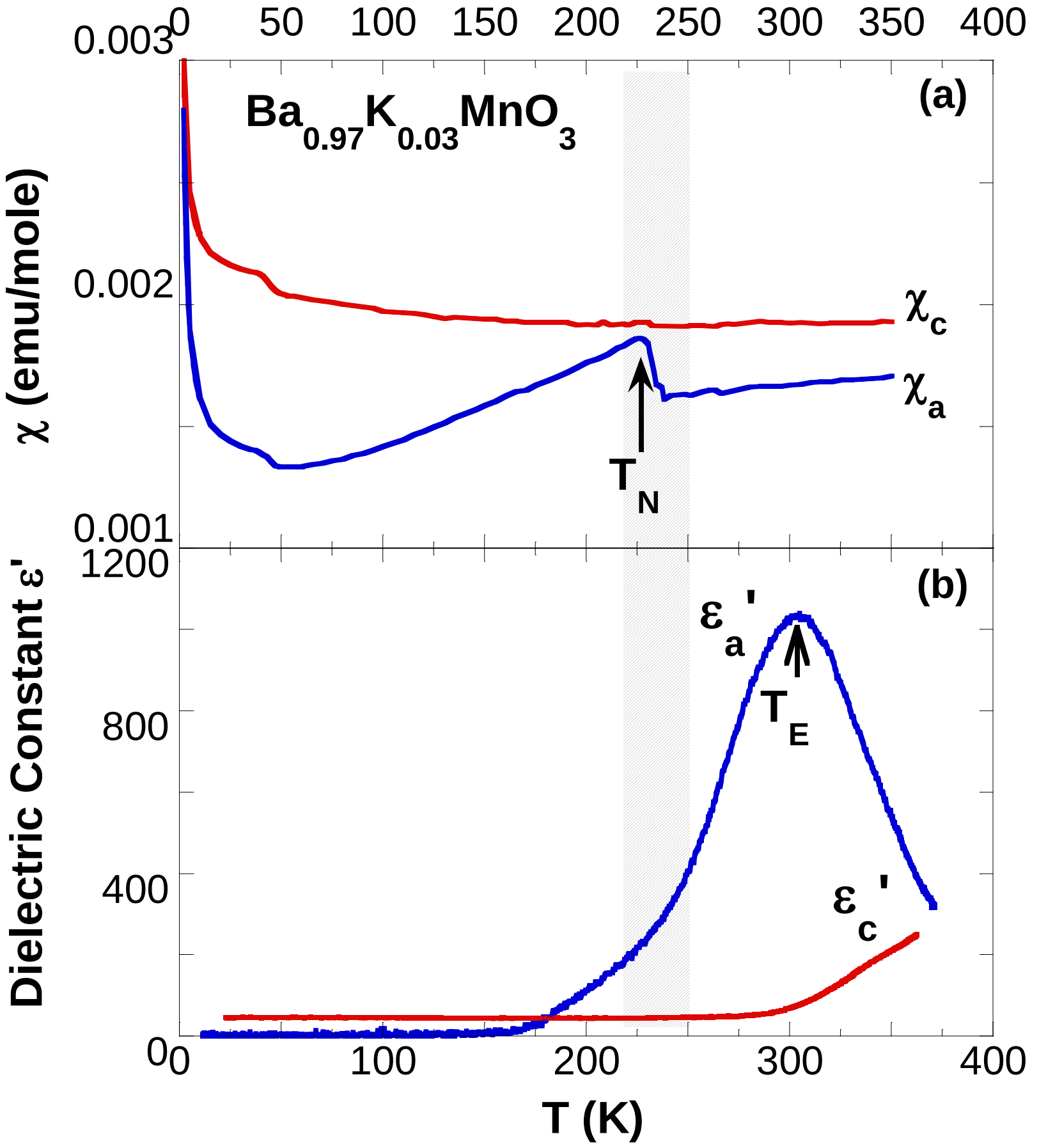

Fig.6

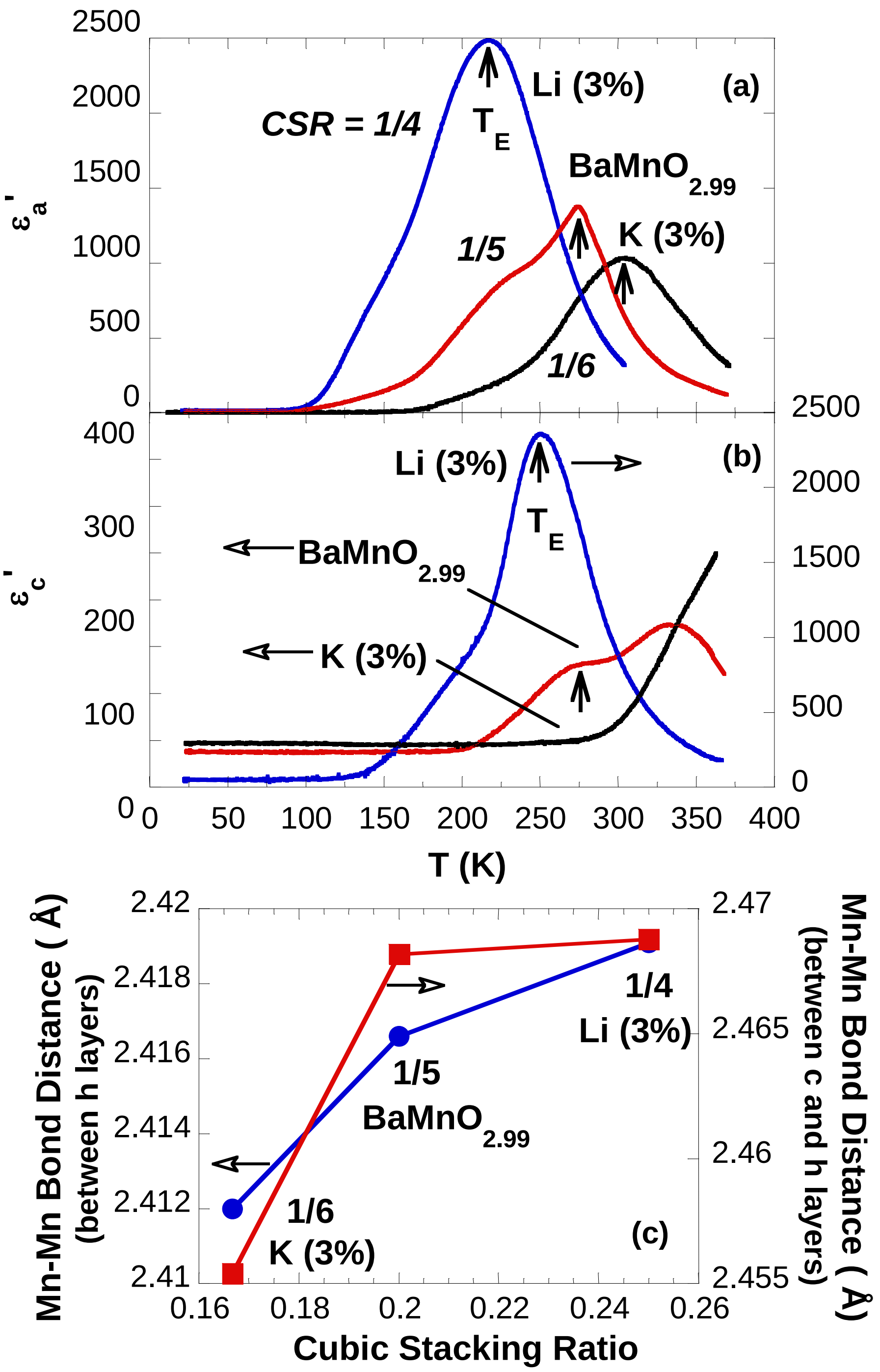

Fig. 7

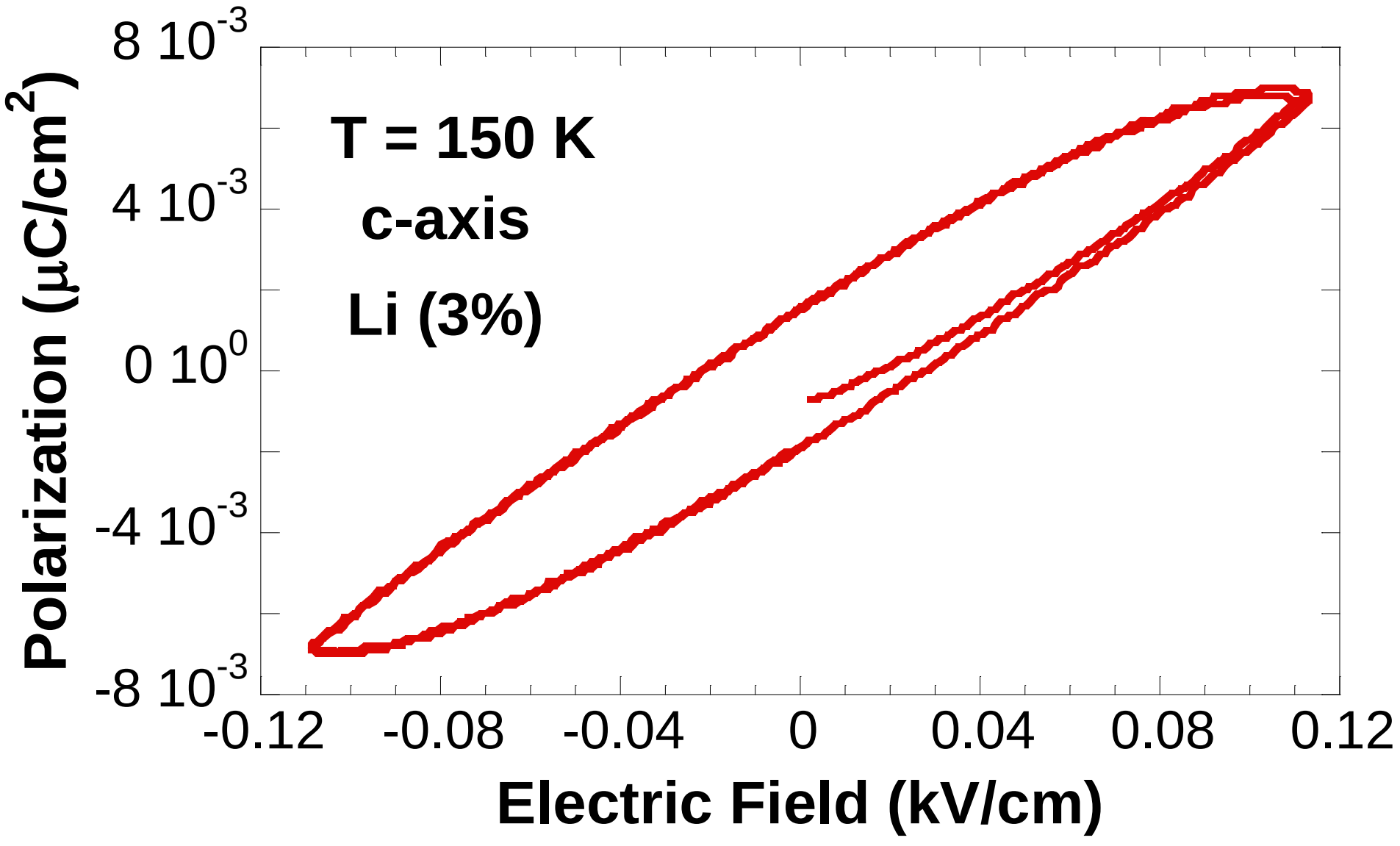

Fig. 8